\documentclass[a4paper]{jpconf}
\pdfoutput=1
\usepackage{graphicx}
\usepackage{amsmath,amsfonts,amsthm}
\usepackage{epsfig,amssymb,wrapfig}
\begin{document}
\title{Properties of the quantum vacuum calculated from its structure}

\author{G. B. Mainland$^1$ and Bernard Mulligan$^2$}

\address{$^1$ Department of Physics, The Ohio State University at Newark, 1179 University Dr., Newark, OH 43055, USA, $^2$ Department of Physics, The Ohio State University, Columbus, OH 43210, USA}

\ead{mainland.1@osu.edu, mulligan.3@osu.edu}

\begin{abstract}
Physicists have speculated about the properties of the quantum vacuum for at least 85 years; however, only recently have they understood the quantum vacuum sufficiently well to begin making testable predictions. Specifically, using Maxwell's equations to describe the interaction of  the electromagnetic field with charged lepton - antilepton vacuum fluctuations, it has been possible to calculate the permittivity of the vacuum, the speed of light in the vacuum, and the fine structure constant.  Physicists are now also beginning to successfully address problems in cosmology based on properties of the quantum vacuum. The terms ``vacuum catastrophe'' and ``old cosmological problem'' refer, respectively, to the predictions that the vacuum energy density  and the cosmological constant are both approximately 120 orders of magnitude larger than the observed values. Using properties of the quantum vacuum and well-established physics, it is possible to demonstrate that the huge vacuum energy cannot transfer energy to normal matter; accordingly, vacuum energy  contributes neither to the observed energy density of the universe nor to the cosmological constant, which plays a central role in the accelerating expansion of the universe.
\end{abstract}

\section{Introduction}
\label{sec:1}
 A central tenet of physics is that the structure of a physical system determines its properties.  Since vacuum fluctuations  manifest and characterize the quantum vacuum, it follows that the electromagnetic properties of the quantum vacuum can be calculated using Maxwell's equations to describe the interaction of an electromagnetic field with vacuum fluctuations.  Electromagnetic properties of the vacuum include the speed $c$ of light in the vacuum, the fine-structure constant $\alpha$, and the permittivity $\epsilon_0$ of vacuum\footnote{The value of  permeability $\mu_0$ of the vacuum (magnetic constant) was originally arbitrarily chosen so that the rationalized meter-kilogram-second unit of current was equal in size to the ampere in the ``electromagnetic (emu)'' system.  As a consequence $\mu_0$, which is a measurement-system constant, not a property of the vacuum, was originally defined to be exactly $\mu_0\equiv 4\pi \times 10^{-7}$ H/m.   As of May 20, 2019,  instead of being defined to have the value  $4\pi \times 10^{-7}$ H/m, $\mu_0\equiv2\alpha h/(ce^2)$, which follows from the definition of the fine-structure constant and $c=1/\sqrt{\mu_0\epsilon_0}$. The change in the definition of $\mu_0$ only affects  the value of $\mu_0$ in about the tenth significant figure and beyond.}.  

 As will be discussed in more detail in the next section, vacuum fluctuations of the fields associated with massive particles\footnote{The generic term ``particle'' will be used to refer to both particles and antiparticles as a category.} appear as particle-antiparticle bound states. Because these bound states oscillate when  interacting with the electromagnetic field associated with a photon, the permittivity of the vacuum can then be calculated using methods somewhat similar to those employed to calculate the permittivity of a dielectric. The resulting approximate formula is
\begin{equation}\label{eqn:1}
\epsilon_0 \cong   \frac{ 6\mu_0}{\pi}\left(\frac{8e^2}{\hbar}\right)^2= 9.10\times 10^{-12}\rm \frac{C}{Vm}\, .
\end{equation}
In the above equation  $\hbar$ is Planck's constant divided by $2\pi$ and $e$ is the magnitude of the renormalized charge of an electron.  The accepted value is $\epsilon_0= 8.85\times 10^{-12}$C/Vm.  The above  formula for $\epsilon_0$ is approximate because it only includes the leading term in a power-series expansion in $\alpha$. 

A formula for the speed $c$ of light in the vacuum can immediately be calculated from the formula for $\epsilon_0$ in \eqref{eqn:1} and $c=1/\sqrt{\mu_0\epsilon_0}$,
\begin{equation}\label{eqn:2}
c \cong   \sqrt{\frac{\pi}{6}}\frac{\hbar}{8e^2\mu_0}= 2.96\times 10^8\rm{m/s}\,.
\end{equation}
To three significant figures the defined value is $c=3.00 \times 10^{8}$m/s. 

In any inertial frame an observer is unable to detect relative motion of the quantum vacuum and the inertial frame; consequently, the observer would conclude that the vacuum and the inertial frame are at rest with respect to each other.  Since the speed of light is determined by the properties of the quantum vacuum and Maxwell's equations, each of which is the same in any inertial frame, the calculated value of the speed of light in the vacuum is the same in every direction in any inertial frame  as required by special relativity.
 
A numerical value for the fine-structure constant $\alpha$ follows immediately  by substituting the formulas  for  $\epsilon_0$ in \eqref{eqn:1} and  $c$ in \eqref{eqn:2} into the defining formula for $\alpha$:
\begin{equation}\label{eqn:3}
\alpha \equiv \frac{e^2}{4 \pi \epsilon_0 \hbar c}\cong \frac{1}{32\sqrt{6 \pi}} = \frac{1}{138.9}\,.
\end{equation}
The experimental value is $\alpha=1/137.04$. It is straightforward to show that if a formula for one of the three quantities $\epsilon_0$, $c$, or $\alpha$ is known, formulas for the other two can immediately be calculated.

Understanding the properties of the quantum vacuum is crucial for understanding cosmology.  The term ``vacuum catastrophe'' refers to the fact that the value of the vacuum energy density of the universe appears to be predicted to be approximately 120 orders of magnitude larger than the observed energy density of the universe. Here it is demonstrated,  using properties of the quantum vacuum and well-established physics, that vacuum energy cannot be converted into normal energy either indirectly through vacuum fluctuations or directly; accordingly, vacuum energy does not contribute to the {\it observed} energy density of the universe. 

The term ``old cosmological problem'' refers to the implication that the cosmological constant, which plays a central role in the accelerating expansion of the universe, must, by virtue of the huge energy density of the universe, be approximately 120 orders of magnitude larger than the observed value.  But since vacuum energy cannot transfer energy to normal energy (and matter), the value of the cosmological constant is unaffected by the vacuum energy density.  

The Inflation Theory of the Universe is based on the assumption of a brief, extreme increase in the size of the universe that occurred shortly after the ``big bang''. Many physicists think that the theory explains why the universe is almost flat (resolving the flatness problem), why the universe is very nearly homogeneous (resolving the horizon problem), and why magnetic monopoles have never been detected (resolving the monopole problem). The specific mechanism that causes inflation is unknown.  The theoretical ideas used here to calculate the speed of light provide a mechanization for inflation: the speed of light in the very early universe would have been much higher than it is now because the much higher temperatures of the early universe would quickly cause bound states of massive particle-antiparticle vacuum fluctuations to disassociate, thereby decreasing the density of bound-state vacuum fluctuations and increasing the speed of light.  The effects of a  higher speed of light in the early universe have been discussed in detail,  but the specific mechanism\cite{Mainland:19,Mainland:20} used by the authors of the present article to calculate the speed of light has not been studied in the context of cosmological inflation.

\section{ Structure  of the quantum vacuum}
\label{sec:2}
\subsection{Vacuum Fluctuations}
\label{subsec:2.1}

Physicists use the term ``vacuum fluctuation'' to describe two very different entities:   Vacuum energy creates type 1 vacuum fluctuations. Field theory provides an explanation for the source of the vacuum energy available for the creation of  type 1 vacuum fluctuations and provides a proof that  type 1 vacuum fluctuations must exist.  In describing the quantum vacuum, there is a free field associated with each known particle; conversely, there is  particle for each free field.  Because  spontaneous symmetry breaking has already occurred, type 1 vacuum fluctuations appear with their physical masses.  Type 1 vacuum fluctuations, which have observable consequences, are present for a time $\Delta t$ as permitted by the Heisenberg uncertainty principle and appear as external particles in a Feynman diagram as shown in Fig. 1. 
\setcounter{figure}{0}
\begin{figure}[h]
\begin{center}
\includegraphics[width=84mm]{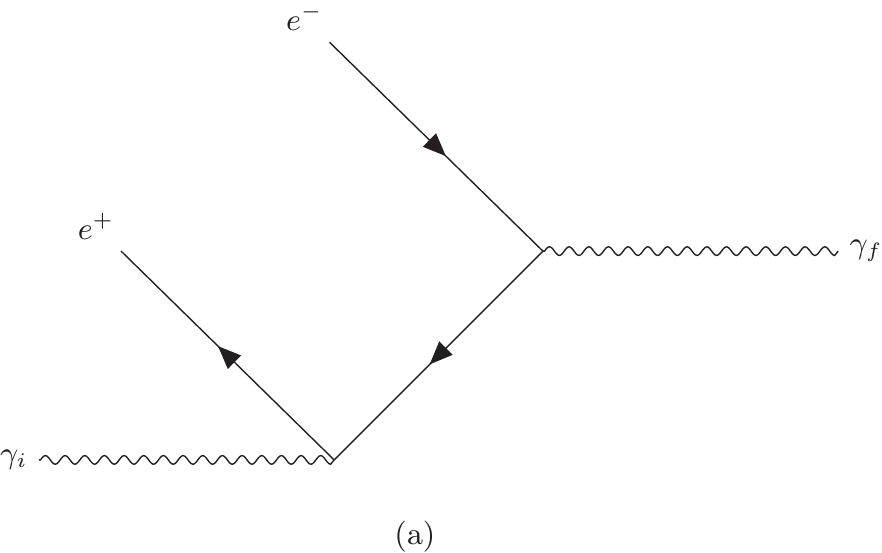}\hspace{0.5 cm}\includegraphics[width=67mm]{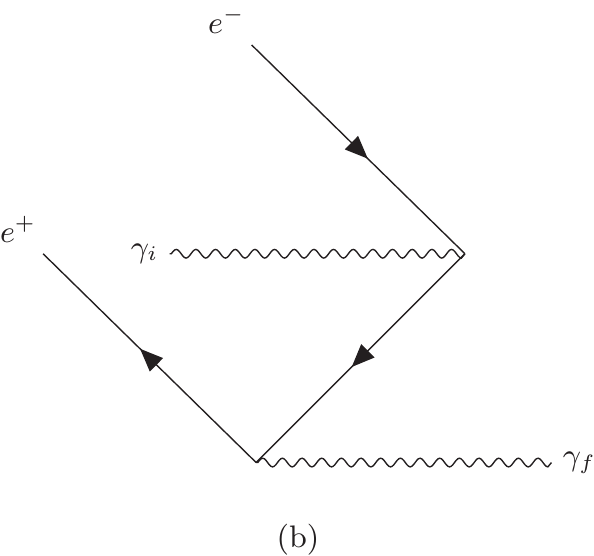}\\
\end{center}
\caption{A parapositronium type 1 vacuum fluctuation consisting of an electron $e^-$ and positron $e^+$ bound in the ground state of positronium is depicted. (a) The incident photon $\gamma_i$ interacts with the positron or (b) the incident photon  interacts with the electron.  In both cases the photon $\gamma_f$, which is identical to $\gamma_i$, is emitted when the parapositronium vacuum fluctuation annihilates.}
\label{fig:1}
\end{figure}
The acronym VF will be used exclusively to refer to  massive, type 1, particle-antiparticle vacuum fluctuations. 

When a VF is created, the laws of physics are obeyed. Consequently, the following quantities are conserved:   angular momentum (the VF itself must have zero angular momentum), electric charge, lepton number and baryon number. The existence of VFs violates energy conservation as permitted by the uncertainty principle.  To minimize the violation of conservation of energy allowed by the Heisenberg uncertainty principle, VFs appear with zero center-of-mass momentum in the least energetic bound state that has zero angular momentum.   Because VFs appear as transient atoms, normal photons interact with VFs somewhat similarly to the way that they interact with normal atoms and molecules. 

VFs are referred to in the literature as being fluctuations of a free field.  Yet earlier in this section it was stated that VFs are associated with fluctuations of fields that have mass.  This seemingly contradictory use of language has its roots in the historical development of field theory. The concept of a  ``free field'' appears in  the text  by Wentzel\cite{Wentzel:03} as early as 1942.  The term  ``field free case'' refers to particles with mass and their associated  fields.  The concept of a free field is treated in the same manner by Roman\cite{Roman:69} where he refers to the ``free field'' as a field that is ``noninteracting''.  It is now understood that the Higgs field, which gives the particle its mass early in the expansion of the universe, is already in place.

Type 2 vacuum fluctuations are  sometimes called vacuum diagrams\cite{Jauch:76} or vacuum bubbles\cite{Bjorken:65,Peskin:95}, a class of Feynman diagrams for which a collection of virtual particles appear from and then vanish back into the vacuum.  However, in contrast to VFs, virtual particles are are off shell and appear only in the interior of Feynman diagrams. Technically a virtual particle is not a particle at all: it appears only in perturbation calculations.   Since Feynman diagrams of vacuum diagrams or vacuum bubbles, and as a consequence type 2 vacuum fluctuations,  do not have any external legs, they cannot contribute to physical processes. They do, however,  contribute to vacuum energy\cite{Peskin:95}.
\begin{figure}[h]
\includegraphics[width=77mm]{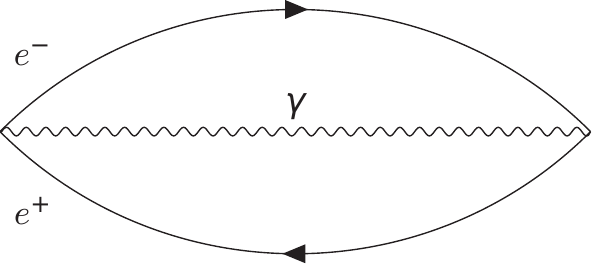}\hspace{0.5 cm}\includegraphics[width=77mm]{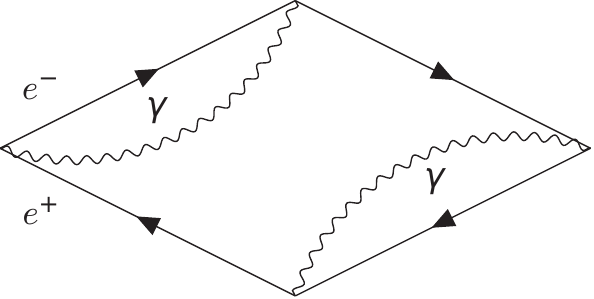}\\
\caption{ Examples of  type 2 vacuum fluctuations that are called vacuum diagrams or vacuum bubbles.}
\label{fig:2}
\end{figure}

\subsection{Poincar\'e invariance of the vacuum}
\label{subsec:2.2}
  
Carroll, Press,  and Turner\cite{Carroll:92} state, ``To a particle physicist, the word `vacuum' has a different meaning than to an astronomer. Rather than denoting `empty space', vacuum is used to mean the ground state (state of lowest energy) of a theory. In general, this ground state must be Lorentz invariant, that is, must look the same to all observers [provided each is in a local inertial frame].'' Since the vacuum is observed through its quantum fluctuations, properties of the vacuum are calculated as expectation values.  The expectation value of each property of the quantum vacuum must be invariant under boosts, space translations and time translations. As a result, the expectation value of each property of the (Minkowski) vacuum is  invariant under  Poincar\'e transformations \cite{Roman:69}, where a Poincar\'e transformation (or inhomogeneous Lorentz transformation) is of the form  \cite{Misner:73},
\begin{equation}\label{eqn:4}
x'^{\mu}=\Lambda^\mu_{\;\; \nu} x^\nu+a^\mu\,.
\end{equation}
Because the expectation value of each property of the vacuum is invariant under time and space translations, to any observer in a local inertial frame the vacuum is homogeneous and time-independent.

If vacuum energy could be ``permanently'' converted into normal energy or vice versa, the  homogeneity of the vacuum would be destroyed in the region where the  transfer of energy between normal and vacuum energy occurred.  When a VF vanishes back into the vacuum, it returns to the vacuum the energy originally  borrowed for its creation, maintaining the homogeneity of the expectation value of the energy density of the vacuum. Vacuum energy is conserved (independently of normal energy), severely restricting the ways in which vacuum energy and normal energy can interact.  The calculation of $\epsilon_0$ discussed in the next section relies crucially on the property that when a VF vanishes back into the vacuum,  the vacuum energy originally required for the creation of the VF is returned to the vacuum.

\section{The interaction of an electron-positron VF with an electromagnetic wave}
\label{sec:3}
\subsection{The number density of electron-positron VFs that are available to interact with a photon}
\label{sec:3.1}

Since the structure of a physical system determines the properties of that system, it is possible to calculate the electromagnetic properties of the vacuum by using Maxwell's equations to describe the interaction of the electromagnetic field of a photon with VFs.  As discussed in the Introduction, if one of the three quantities $\epsilon_0$, $c$, or $\alpha$ has been calculated, then formulas for the other two are easily obtained. Among the three quantities, the possibility of calculating the fine-structure constant has garnered the most attention.  In 1985 Richard Feynman\cite{Feynman:85} wrote about the fine-structure constant, ``It has been a mystery ever since it was discovered more than fifty years ago, and all good theoretical physicists put this number up on their wall and worry about it.$\dots$  It's one of the {\it greatest} damn mysteries of physics: a {\it magic number} that comes to us with no understanding by man. You might say the `hand of God' wrote that number, and `we don't know how He pushed his pencil.'''  In his letter\cite{Miller:94} to Heisenberg dated 14 June 1934, Pauli wrote, ``I have been musing over the great question, what is $e^2/\hbar c$ [$e^2/4 \pi \epsilon_0\hbar c$ in SI units]?'' In response to this statement, Arthur I. Miller, Professor of History and Philosophy of Science, writes, ``We recall that, fundamental to the thinking of Heisenberg and Pauli, was that a theory which can fix the value of $e^2/(4\pi \epsilon_0 \hbar c)$ most likely will have no divergent quantities.''\cite{Miller:94}

Because a method for calculating the permittivity of a dielectric medium has been known for almost a century\cite{Kramers:24}, it follows that first calculating the permittivity of the vacuum, instead of $c$ or $\alpha$, is almost certainly the approach most likely to succeed.  Charged lepton-antilepton VFs, each of which appears at rest in the vacuum in the most tightly bound state that has zero angular momentum,  are the primary contributors to the permittivity of the vacuum. The contribution to $\epsilon_0$ from electron-positron VFs, which appear in the vacuum as parapositronium, is calculated first.  The contribution from muon-antimuon VFs and tau-antitau VFs can then be calculated immediately by replacing the electron mass with the mass of the muon or tau, respectively.
  
Even in an intense laser beam the number density of photons is much less than the number density of VFs\cite{Mainland:19,Mainland:20}.  Thus, the vast majority of parapositronium VFs appear and then disappear without interacting.  Those few that do interact with a photon almost always interact with only one.  When a photon interacts with a parapositronium VF, the electric field associated with the  photon polarizes the the VF.  In the presence of an electromagnetic wave with the  $x$-axis chosen to point in the direction of the electric field of the wave, the induced dipole moment of a VF is  $p_x=e(x_+-x_- )\equiv ex$, where $x_+$ and $x_-$ are, respectively, the coordinates of the  positron and electron.  The potential energy $H_{\rm dipole}$ of the electric dipole of the VF in the electric field $E_x=E_0\cos\omega t$ is
\begin{equation}\label{eqn:5}
H_{\rm dipole}=-p_x\,E_x=-exE_0\cos \omega t\,.
\end{equation}
If a parapositronium VF interacted with an electric field consisting of many photons, the VF would experience a force oscillating at the angular frequency $\omega$. However, since the parapositronium VF almost always interacts with only one photon, which is absorbed at the time $t_i$ when the photon and parapositronium VF interact, the electric field that interacts with the VF is $E(t_i) = E_0\cos \omega t_i \equiv \mathbb{E}_0$. Consequently, to describe the interaction of a single photon with a parapositronium VF,  \eqref{eqn:5} becomes
\begin{equation}\label{eqn:6}
H^{\rm VF}_{\rm dipole}=-exE_0\cos \omega t_i \equiv -ex\mathbb{E}_0\,,
\end{equation}
implying that the VF experiences an impulse. 

The electric displacement $D(t)$\cite{Jackson:99} in a dielectric satisfies 
\begin{equation}\label{eqn:7}
D(t)=\epsilon E(t)=\epsilon_0 E(t) +P(t)\,,
\end{equation}
where $\epsilon$ is the permittivity of the dielectric and $E(t)$ is the electric field. As discussed earlier, the interaction of a photon with a vacuum fluctuation occurs one-at-a-time. Since each interaction occurs in exactly the same manner, it is appropriate to describe these interactions as a property of the vacuum as a whole. Thus \eqref{eqn:7} properly describes the photon-VF interaction as a property of the vacuum with a polarization density $P(t)$ given by
\begin{equation}\label{eqn:8}
P(t)=\sum_j N_j \langle p_j(t)\rangle\,.
\end{equation}
In \eqref{eqn:8}   $N_j$ is the number of oscillators of the $j^{\rm th}$ variety per unit volume that  are available to interact with a photon, and $\langle p_j(t)\rangle $ is the expectation value of the dipole moment of the $j^{\rm th}$ variety. From \eqref{eqn:7} it follows that the increase in $D(t)$ from the value $\epsilon_0 E(t)$ in the vacuum to its value  $\epsilon E(t)$ in the dielectric, results entirely from the polarization density $P(t)$ of the dielectric.  In the vacuum it then follows that  $\epsilon_0E(t_i)$ results entirely from the polarization density $P^{\rm VF}(t_i)$ of VFs,
\begin{equation}\label{eqn:9}
\epsilon_0  E(t_i)= P^{\rm VF}(t_i)=\sum_j N^{\rm VF}_j \langle p^{\rm VF}_j(t_i)\rangle\,,
\end{equation}
where the final equality follows from  \eqref{eqn:8}.  As will be shown, the right-hand side of \eqref{eqn:9} is proportional to the electric field $E(t_i)$ that polarizes the VF; therefore, the electric field cancels out of the equation, yielding an equation for $\epsilon_0$. The calculation of $\epsilon_0$ is carried out below by first calculating $N^{\rm VF}_j$  for a parapositronium VF and then calculating the expectation value $\langle p^{\rm VF}_j(t_i)\rangle$ of the dipole moment  induced in a parapositronium VF by the electric field $E(t_i)$  of a photon. The complication in calculating the polarization density of vacuum fluctuations instead of normal matter arises because a vacuum fluctuation has a finite lifetime while an atom or molecule in a typical dielectric is stable.

The crucial characteristic of the atoms and molecules that form dielectrics is that they oscillate when interacting with photons or an electric field. Consequently, it is convenient to describe a parapositronium atom as an oscillator. Feynman\cite{Feynman:64} writes that when an atom is in its ground state, the ionization energy $E_{\rm ionization} =\hbar \omega^0$, where $\omega^0$ is the angular frequency of the harmonic oscillator that corresponds to the oscillator.  Feynman's condition is equivalent to requiring that  the expectation value of $x^2$ for a parapositronium atom equals the expectation value of $x^2$ for the corresponding oscillator.
 
The binding (ground-state) energy of parapositronium is obtained from the ground-state energy of hydrogen by replacing the reduced mass of hydrogen with the reduced mass of parapositronium. In this replacement $\mu_{\rm hydrogen}$, which is approximately $m_e, \longrightarrow \mu_{\rm p-Ps} = m_e/2$, where $m_e$ is the mass of an electron. Accordingly,
\begin{equation}\label{eqn:10}
E_{\rm p-Ps} = -\frac{ (m_e/2)e^4 }{2(4\pi \epsilon_0)^2 \hbar^2} =-\frac{m_e\alpha^2c^2}{4} \,. 
\end{equation}
The corresponding angular frequency $\omega_{\rm p-Ps}^0$ of parapositronium is
\begin{equation}\label{eqn:11}
\omega_{\rm p-Ps}^0=\frac{|E_{\rm p-Ps}|}{\hbar} = \frac{m_e\alpha^2c^2}{4\hbar} \,. 
\end{equation}

To minimize the violation of conservation of energy when a VF appears in the vacuum, the center of mass of a VF is at rest. The zitterbewegung of the VF gives the VF its size\cite{Gersch:92,Corinaldesi:63},  
\begin{equation}\label{eqn:12}
L_{\rm p-Ps}\equiv  \frac{\hbar}{4m_ec} \,.
\end{equation}
Requiring that there be only one VF in the  volume $(L_{\rm p-Ps})^3$, the number of parapositronium VFs per unit volume\footnote{Without comment, in formula (1X.1) Zel'dovich\cite{Zeldovich:68} uses a value for  a length corresponding to \eqref{eqn:12} that is twice as large, yielding a number density that is one eighth as large.} is $1/(L_{\rm p-Ps})^3$.

The Heisenberg uncertainty principle relating the uncertainty in time $ \Delta t$ and the uncertainty in energy $\Delta E$ is 
\begin{equation}\label{eqn:13}
\Delta E\, \Delta t\geq\frac{\hbar} {2} \, .
\end{equation}
 The lifetime $\Delta t_{\rm p-Ps}$ of a parapositronium  VF is the minimum time $\Delta t$ in \eqref{eqn:13}.   During the lifetime $\Delta t_{\rm p-Ps}$, the violation of conservation of energy   $\Delta E_{\rm p-Ps}$ is the total energy of the parapositronium VF, which is the energy $2m_ec^2$ associated with the masses of the electron and positron minus the binding energy of parapositronium given in \eqref{eqn:10}. Neglecting the binding energy because it is very small in comparison with  $2m_ec^2$, $\Delta E_{\rm p-Ps}\cong 2m_ec^2$, and \eqref{eqn:13} yields 
\begin{equation}\label{eqn:14}
\Delta t_{\rm p-Ps}\cong \frac{\hbar}{4m_ec^2} \, .
\end{equation}

The decay rate of a photon-excited parapositronium VF is\cite{Mainland:20}
\begin{equation}\label{eqn:15}
\Gamma_{\rm p-Ps} = \frac{\alpha^5 m_e c^2}{ \hbar} \, .
\end{equation}

The lifetime $\Delta t_{\rm p-Ps}$ of a parapositronium VF is $\cong 3.2 \times10^{-22}$s whereas the lifetime of  of  a photon-excited (polarized) parapositronium VF is $1/\Gamma_{\rm p-Ps}\cong 6.2 \times 10^{-11}$s, which is approximately $10^{11}$ times as long. Because a photon-excited parapositronium VF has a much greater lifetime that a  parapositronium VF, and, therefore, is much more stable, the former is called a quasi-stationary state\cite{Davydov:76}.

The probability that the photon-excited parapositronium VF has not decayed after a time $t$ is $e^{-\Gamma_{\rm p-Ps}t}$, and  the probability that it has  decayed is $1-e^{-\Gamma_{\rm p-Ps}t}$. For each photon-VF interaction, the probability for a parapositronium VF to  interact with a photon and form a polarized parapositronium VF  equals the probability for the  polarized parapositronium VF to  annihilate and emit a photon; consequently, during the lifetime $\Delta t_{\rm p-Ps}$ the average probability of a parapositronium VF interacting with a photon is $1-e^{-\Gamma_{\rm p-Ps}\Delta t_{\rm p-Ps}}$.

For a parapositronium VF the quantity $N^{\rm VF}_j \equiv N^{\rm VF}_{\rm p-Ps}$ in \eqref{eqn:9} is the product of the number of parapositronium VFs per unit volume, $1/(L_{\rm p-Ps})^3$, multiplied by the probability $1-e^{-\Gamma_{\rm p-Ps}\Delta t_{\rm p-Ps}}$ of a parapositronium VF interacting with a photon during the lifetime of the parapositronium VF,
\begin{align}\label{eqn:16}
N^{\rm VF}_{\rm p-Ps}&\cong \frac{1}{(L_{\rm p-Ps})^3}\times(1- e^{-\Gamma_{\rm p-Ps}\, \Delta t_{\rm p-Ps}}) \,.
\end{align}
Noting that $\Gamma_{\rm p-Ps}\,\Delta t_{\rm p-Ps}\ll1$, expanding the above exponential in a Maclaurin series, and keeping only the first two terms,
\begin{align}\label{eqn:17}
N^{\rm VF}_{\rm p-Ps} \cong \frac{1}{(L_{\rm p-Ps})^3}\times \Gamma_{\rm p-Ps} \; \Delta t_{\rm p-Ps}=  \frac{\alpha^5}{4}{\left ({\frac{4 m_e c}{\hbar}} \right )}^3\,,
\end{align}
where \eqref{eqn:12}, \eqref{eqn:14}, and \eqref{eqn:15} were used to obtain the final equality.

\subsection{Expectation value of the electric dipole moment induced in a electron-positron (parapositronium) VF}
\label{sec:3.2}

\noindent As discussed in Sec.~3.1, the oscillator corresponding to a parapositronium VF has the resonant frequency $\omega^0_{\rm p-Ps}$ given in \eqref{eqn:11}. Using the oscillator model of a parapositronium VF,  in this section the expectation value $\langle p^{\rm VF}_{\rm p-Ps}\rangle$ is calculated for the electric dipole moment induced in a parapositronium VF by an electric field $E(t_i)\equiv \mathbb{E}_0$ associated with a photon.

The  Schr\"odinger equation for a one-dimensional harmonic oscillator  is 
\begin{equation}\label{eqn:18}
H_{\rm osc} \psi(x)=\left [-\frac{\hbar^2}{2\mu}\frac{{\rm d}^2}{{\rm d}x^2}+\frac{1}{2}\mu (\omega^0)^2x^2\right ]\psi(x)=E\psi(x)\,, 
\end{equation}
where, as discussed in Sec.~3.1, $x$ is the relative position of the  positron and electron in the parapositronium VF. The normalized,  ground-state solution\cite{Park:05} to \eqref{eqn:18} is
\begin{equation}\label{eqn:19}
\psi^0(x)=\left (\frac{\mu \omega^0}{\pi \hbar}\right )^\frac{1}{4}e^{-\mu \omega^0x^2/2\hbar}\,.\\
\end{equation}

The Hamiltonian describing a parapositronium VF that has been polarized by interacting with the electric field of a photon is the sum of the oscillator Hamiltonian $H_{\rm osc}$ in \eqref{eqn:18} and the dipole Hamiltonian $H^{\rm VF}_{\rm dipole}$ in \eqref{eqn:6}. Thus the  Schr\"odinger equation is
\begin{equation}\label{eqn:20}
\left[-\frac{\hbar^2}{2\mu_{\rm p-Ps}}\frac{{\rm d}^2}{{\rm d}x^2}+\frac{1}{2}\mu_{\rm p-Ps} (\omega_{\rm p-Ps}^0)^2x^2-ex\mathbb{E}_0\right ]\psi_{\rm p-Ps}(x)=E\psi_{\rm p-Ps}(x)\,.
\end{equation}
Factoring $\frac{1}{2}\mu_{\rm p-Ps} (\omega^0_{\rm p-Ps})^2$ out of the two potential energy terms, \eqref{eqn:20} becomes
\begin{equation}\label{eqn:21}
\left \{ -\frac{\hbar^2}{2\mu_{\rm p-Ps}}\frac{{\rm d}^2}{{\rm d}x^2}+\frac{1}{2}\mu_{\rm p-Ps} (\omega_{\rm p-Ps}^0)^2\left [x^2-\frac{2e\mathbb{E}_0}{\mu_{\rm p-Ps} (\omega^0_{\rm p-Ps})^2}x\right ]\right 
\}\psi_{\rm p-Ps}(x)=E\psi_{\rm p-Ps}(x)\,.
\end{equation}
The above equation can be solved exactly by ``completing the square'' of the two terms in square brackets.  When written in terms of the coordinate $u$ defined by
\begin{equation}\label{eqn:22}
u\equiv x-\frac{e\mathbb{E}_0}{\mu_{\rm p-Ps}(\omega_{\rm p-Ps}^0)^2}\,,
\end{equation}
\eqref{eqn:21} takes the desired form:
\begin{equation}\label{eqn:23}
\left[-\frac{\hbar^2}{2\mu_{\rm p-Ps}}\frac{{\rm d}^2}{{\rm d}u^2}+\frac{1}{2}\mu_{\rm p-Ps} (\omega_{\rm p-Ps}^0)^2u^2-\frac{(e\mathbb{E}_0)^2}{2\mu_{\rm p-Ps}(\omega_{\rm p-Ps}^0)^2}\right ] \psi_{\rm p-Ps}(u)=E\psi_{\rm p-Ps}(u)\,.
\end{equation}
The first two terms on the left-hand side of \eqref{eqn:23} are the Hamiltonian of a harmonic oscillator; therefore, in comparison with an unpolarized oscillator described by the Hamiltonian $H_{\rm osc}$ in  \eqref{eqn:18}, each energy level  in the polarized oscillator described by the Schr\"odinger equation \eqref{eqn:23} is lowered by $(e\mathbb{E}_0)^2/[2\mu_{\rm p-Ps}(\omega_{\rm p-Ps}^0)^2]$, the change in the energy of a positronium VF in the presence of a polarizing photon.

From \eqref{eqn:19} the exact, normalized, ground-state solution to \eqref{eqn:23} is
\begin{equation}\label{eqn:24}
\psi^0_{\rm p-Ps}(u)=\left (\frac{\mu_{\rm p-Ps}\, \omega_{\rm p-Ps}^0}{\pi \hbar}\right )^\frac{1}{4}e^{-\mu_{\rm p-Ps}\, \omega_{\rm p-Ps}^0\,u^2/2\hbar}\,.\\
\end{equation}
The expectation value $\langle p^{\rm VF}_{\rm p-Ps}\rangle$ of the electric dipole moment of a polarized parpositronium VF in the state characterized by $\psi^0_{\rm p-Ps}(u)$  is
\begin{equation}\label{eqn:25}
\langle p^{\rm VF}_{\rm p-Ps}\rangle=\int_{-\infty}^\infty{\rm d}u\,\psi^{0*}_{\rm p-Ps}(u)\,ex\,\psi^0_{\rm p-Ps}(u)=\frac{e^2\mathbb{E}_0} {\mu_{\rm p-Ps}(\omega_{\rm p-Ps}^0)^2}\,. 
\end{equation}
Using $\mu_{\rm p-Ps}=m_e/2$ and the explicit expression for $\omega_{\rm p-Ps}^0$ in \eqref{eqn:11},
\begin{equation}\label{eqn:26}
\langle p_{\rm p-Ps}^{\rm VF}\rangle=\frac{32 \hbar^2e^2}{\alpha^4 c^4 m_e^3}\mathbb{E}_0 \, .
\end{equation}

Combining the above equation and  \eqref{eqn:17}, for parapositronium the right-hand side  of  \eqref{eqn:9} becomes
\begin{equation}\label{eqn:27}
N_{\rm p-Ps}^{\rm VF}\langle p_{\rm p-Ps}^{\rm VF}\rangle=\frac{8^3 \alpha e^2}{\hbar c}\mathbb{E}_0\,.
\end{equation}
Since the mass has cancelled in the above equation, in  \eqref{eqn:9} the contributions from muon-antimuon VFs and tau-antitau VFs are the same as that from electron-positron VFs.  As a result   \eqref{eqn:9} becomes
\begin{equation}\label{eqn:28}
\epsilon_0  \mathbb{E}_0 = \sum_{j=1}^3\frac{8^3 \alpha e^2}{\hbar c}\mathbb{E}_0=\frac{3(8^3) \alpha e^2}{\hbar c}\mathbb{E}_0\,.
 \end{equation}
The electric field $E(t_i) \equiv \mathbb{E}_0$ cancels, yielding the  formula
\begin{equation}\label{eqn:29}
\epsilon_0 = \frac{3(8^3) \alpha e^2}{\hbar c}= \frac{ 6\mu_0}{\pi}\left(\frac{8e^2}{\hbar}\right)^2\,.
 \end{equation}
 The final equality in \eqref{eqn:29} Is obtained using the definition of $\alpha$ in  \eqref{eqn:3} and the formula $c=1/\sqrt{\mu_0\epsilon_0}$.

Quark-antiquark VFs also form bound states that can oscillate when interacting with the electric field associated with a photon. The maximum contribution to $\epsilon_0$ from any of these VFs  is estimated to be smaller than the contribution from charged lepton-antilepton VFs by a factor of about $10^{-4}$\cite{Mainland:19}.

\section{ Resolution of the ``vacuum catastrophe'' and the ``old cosmological constant problem''}
\label{sec:4}

The  ``vacuum catastrophe'' is a calculation of the  vacuum energy density of the universe that is approximately 120 orders of magnitude larger than the observed energy density of the universe.  In this section an explanation is given for why the calculation leading to the ``vacuum catastrophe'' is flawed. When calculated correctly, there is no ``vacuum catastrophe''. The ``old cosmological constant problem'' occurs because the calculated value of the cosmological constant is approximately 120 orders of magnitude larger than the observed  value.  The ``vacuum catastrophe'' and the ``old cosmological constant problem'' have the same origin: once it is understood that there is no ``vacuum catastrophe'', it follows immediately that there is also no ``old cosmological constant problem''.

In Sec. 2.2 it was pointed out that Poincar\'e invariance of the vacuum implies that vacuum energy is conserved. The action integral describing the interaction of normal energy (and matter) is invariant under time translations.  From Noether's theorem\cite{Noether:18}, it then follows that normal energy is conserved (independently of vacuum energy).  In curved spacetime, however,  Birrell and Davies\cite{Birrell:94} point out that, ``$\dots$the Poincar\'e group is no longer a symmetry group of the spacetime.'' Therefore, when using general relativity to describe normal energy (and matter), a central question is, ``Are vacuum energy and normal energy still conserved independently?'' The action integral for general relativity is invariant under time translations; consequently, from  Noether's theorem\cite{Noether:18}, normal energy is still conserved. The requirement that the vacuum be the same in any inertial frame implies that  the vacuum is homogeneous and is described by a Minkowski space. As before, vacuum energy is conserved independently of normal energy. Even in the presence of a huge ``normal'' gravitational field, the vacuum is not described by curved spacetime (a Riemannian manifold): the ``normal'' gravitational field cannot do net work on vacuum energy (or matter) with the result that the spacetime of the  vacuum does not become curved.

The ``vacuum catastrophe'' occurs when it is assumed that vacuum energy exerts forces on normal  energy (and matter) in the same way that normal energy exerts forces on other normal  energy (and matter). The assumption is false. Vacuum fluctuations are the manifestation of vacuum energy.  During the existence of a vacuum fluctuation,  it cannot do net work on normal  matter and energy because the energy associated with that work would violate the independent conservation of  vacuum energy and normal energy.  Also, if vacuum energy could create a normal gravitational field, that field could do work on normal energy, violating the independent conservation of normal energy and vacuum energy.  Since normal energy is conserved independently of vacuum energy, normal energy  cannot do net work on a vacuum fluctuation during its existence. The ``vacuum catastrophe'' does not actually exist because vacuum energy cannot do net work on normal energy and vice versa: it is incorrect to treat vacuum energy as if it were normal energy. 

The first calculations of the the Casimir effect\cite{Casimir:48a,Casimir:48b} were based on the idea that the Casimir force was caused by vacuum fluctuations of the electromagnetic field.  If this were so, the Casimir force, which acts on normal matter, would convert vacuum energy into normal energy so that the independent conservation of normal energy and vacuum energy would be violated.  In more recent articles\cite{Schwinger:75,Schwinger:78,Milonni:82,Watanabe:98} the Casimir effect has been explained with vacuum energy playing no role.

The Planck CMB\cite{Planck:18} anisotropy measurements established that the energy density in the universe is, to within a fraction of a percent,  equal to the critical energy density\cite{Tanabashi:18},
\begin{equation}\label{eqn:30}
\rho^{\rm energy}_{\rm universe}\cong \rho^{\rm energy}_{\rm critical}=\frac{3(H_0 c)^2}{8 \pi G}=7.76\times 10^{-10}{\rm J/m^3}\,,
\end{equation}
where $G$ and $H_0$ are, respectively, the gravitational force constant and the present value of the Hubble constant.  Imposing the condition that the maximum energy of a photon is the Planck energy $E_p=\sqrt{\hbar c^5/G}$, the vacuum energy density for photons\cite{Adler:95} is, 
\begin{equation}\label{eqn:31}
\rho^{\rm vacuum\; energy}_{\rm photons}=\frac{c^7}{8\pi^2 \hbar G^2} =5.87 \times 10^{111}\,\mbox{J/m}^3 \,.
\end{equation}
Comparing the above two equations, the observed energy density of the universe is about $10^{120}$ times smaller than the theoretical energy density resulting from just from photons. Including the vacuum energy density from the  the eight gluons and the graviton will increase the theoretical vacuum energy density by an order of magnitude or so. The contribution from fundamental, massive particles is much less\cite{Mainland:20}.

When a cosmological constant $\Lambda_{\rm}$  is included,  Einstein's field equation\cite{Misner:73} is
\begin{equation}\label{eqn:32}
R_{\alpha \beta}-\frac{1}{2}R\, g_{\alpha \beta}+\Lambda \, g_{\alpha \beta}=\frac{8\pi G}{c^4}
T_{\alpha \beta}\,.
\end{equation}
In \eqref{eqn:32} $R_{\alpha \beta}$,  $R \equiv R^\mu_\mu$, $g_{\alpha \beta}$,  and   $T_{\alpha \beta}$ are, respectively, the Ricci tensor, the Ricci scalar, the metric tensor,  and  the energy-momentum tensor.  

To understand how the ``old cosmological constant problem'' arises, the  energy-momentum tensor is split into two parts: that arising from normal energy (and matter) and that from vacuum energy (and matter). The incorrect assumption is then made that vacuum energy  interacts with normal energy in the same way that normal energy  interacts with other normal energy. 
\begin{equation}\label{eqn:33}
T_{\alpha \beta}=T^{\rm normal\; energy}_{\alpha \beta}+T^{\rm vacuum \; energy}_{\alpha \beta}\,.
\end{equation}
Substituting \eqref{eqn:33} into \eqref{eqn:32},
\begin{equation}\label{eqn:34}
R_{\alpha \beta}-\frac{1}{2}R\, g_{\alpha \beta}+\Lambda\; g_{\alpha \beta} -\frac{8\pi G}{c^4}T_{\alpha \beta}^{\rm vacuum \;energy}=\frac{8\pi G}{c^4}T_{\alpha \beta}^{\rm normal \;energy}\,.
\end{equation}

As will now be shown, the fact that the vacuum is the same in any inertial reference frame forces the energy-momentum tensor of the vacuum to have the form of a cosmological constant.  First consider $T_{\alpha \beta}^{\rm vacuum \;energy}$ in Minkowski space.  The only tensor that is invariant under under Lorentz boosts is the metric tensor $\eta_{\alpha \beta}$\cite{Carroll:92}, which is diagonal with elements chosen here to be (-1,1,1,1). As a result,
\begin{equation}\label{eqn:35}
T_{\alpha \beta}^{\rm vacuum \;energy}={\cal S} \eta_{\alpha \beta}\,,
\end{equation}
where ${\cal S}$ is a scalar.

The vacuum possesses the properties of a perfect fluid: it cannot conduct heat,  exert shear stress or possess viscosity. Modeling the vacuum as a perfect fluid, the energy-momentum tensor of the vacuum is that of a perfect fluid\cite{Misner:73}.
\begin{equation}\label{eqn:36}
 T_{\alpha \beta}^{\rm vacuum \;energy} = \left(\rho^{\rm vacuum \;energy}+P^{\rm vacuum} \right )\frac{1}{c^2}U_\alpha U_\beta +P^{\rm vacuum}\, \eta_{\alpha \beta}\,,
\end{equation}
where  $\rho^{\rm vacuum \;energy}$,  $P^{\rm vacuum}$, and $U_\alpha$ are, respectively, the energy density, isotropic pressure, and 4-velocity of the vacuum.  In any inertial frame the vacuum is at rest, implying that the three-velocity $\mathbf{v}$ of the vacuum is zero. Using $U_\alpha(\mathbf{v}=0)=(c,0,0,0)$  the vacuum has the diagonal energy-momentum tensor $T_{\alpha \beta}$ where
\begin{equation}\label{eqn:37}
T^{\rm vacuum \;energy}_{00}=\rho^{\rm vacuum \;energy},\;T^{\rm vacuum \;energy}_{11}=T^{\rm vacuum \;energy}_{22}=T^{\rm vacuum \;energy}_{33}=P^{\rm vacuum}\,.
\end{equation}
Using \eqref{eqn:37},  \eqref{eqn:33} is satisfied provided
\begin{subequations}\label{eqn:38}
\begin{align}
\label{eqn:38a}
&{\cal S}= -\rho^{\rm vacuum \; energy}\,,\\
\label{eqn:38b}
&P^{\rm vacuum}=-\rho^{\rm vacuum \; energy}\,.
\end{align}
\end{subequations}
Eq. \eqref{eqn:38b} is the equation of state for the vacuum, and the negative pressure in that equation drives the expansion of the universe. Combining \eqref{eqn:38a} and \eqref{eqn:35},
\begin{equation}\label{eqn:39}
T_{\alpha \beta}^{\rm vacuum \;energy}=-\rho^{\rm vacuum \; energy}\, \eta_{\alpha \beta}\,.
\end{equation}
In curved space-time \eqref{eqn:39} would become
\begin{equation}\label{eqn:40}
T_{\alpha \beta}^{\rm vacuum \;energy}=-\rho^{\rm vacuum \;energy}\; g_{\alpha \beta}\,.
\end{equation}
Using  \eqref{eqn:40}, \eqref{eqn:34} would take the form,  
\begin{equation}\label{eqn:41}
R_{\alpha \beta}-\frac{1}{2}R\, g_{\alpha \beta}+\left (\Lambda+\frac{8\pi G}{c^4}\rho^{\rm vacuum \; energy}\right )g_{\alpha \beta}=\frac{8\pi G}{c^4}T_{\alpha \beta}^{\rm normal \;energy}\,.
\end{equation}
Eq. \eqref{eqn:41} reveals that if vacuum energy interacts as if it were normal energy, then vacuum energy would induce a cosmological constant with a value $(8\pi G/c^4) \rho^{\rm vacuum \;energy}$; consequently, the effective cosmological constant $\Lambda_{\rm eff}$ would be
\begin{equation}\label{eqn:42}
\Lambda_{\rm eff}= \Lambda+ \frac{8\pi G}{c^4}\rho^{\rm  vacuum \; energy}\,.
\end{equation}

The ``old cosmological constant problem''  arises from \eqref{eqn:42}.  An order-of-magnitude estimate of the last term in \eqref{eqn:42} is obtained by approximating (but underestimating)  the vacuum energy density by the vacuum energy density of photons given in \eqref{eqn:31}, 
\begin{equation}\label{eqn:43}
 \frac{8\pi G}{c^4}\rho^{\rm  vacuum \; energy} \stackrel{>}{\approx }\frac{8\pi G}{c^4}\rho^{\rm  vacuum \; energy}_{\rm photons}  = 10^{69}{\rm m}^{-2}\,.
\end{equation}
The experimental value of the cosmological constant is obtained from $\Lambda= 8 \pi G \rho^{\rm mass}_{\rm critical}\Omega_\Lambda/c^2$, where $\Omega_\Lambda =0.692$\cite{Tanabashi:18}, and the critical mass density of the universe  $\rho^{\rm mass}_{\rm  critical}= 8.63\times 10^{-27}$ kg/m$^3$\cite{Tanabashi:18},
\begin{equation}\label{eqn:44}
 \Lambda_{\rm expt} = 1.11 \times10^{-52} {\rm m}^{-2}\, .
\end{equation}
 The ``old cosmological problem'' is the fact that the experimental value $\Lambda_{\rm expt}$  appears to be approximately $10^{-121}$ times smaller than the  underestimate of the contribution to $\Lambda$ from the vacuum energy density in \eqref{eqn:43}.  Both the ``vacuum catastrophe'' and  the ``old cosmological problem'' are caused by incorrectly assuming that vacuum energy can exert a ``permanent''  gravitational force on normal matter, which it cannot. Since vacuum energy cannot exert gravitational forces, it is incorrect to include the energy-momentum tensor of the vacuum in Einstein's field equation \eqref{eqn:41}, eliminating the contribution to the cosmological constant  from vacuum energy in \eqref{eqn:42}.
 
When the universe is modeled as a perfect fluid, the vacuum energy density remains constant as the universe expands.  As is customary, the pressure is taken to be inward.  When the vacuum expands by an infinitesimal volume $\Delta V$, the work $\Delta W$ done by the pressure $P^{\rm vacuum}$ is $\Delta W= -P^{\rm vacuum}\,\Delta V$. Using  \eqref{eqn:38b},
\begin{equation}\label{eqn:45}
\Delta W= \rho^{\rm vacuum\, energy}\,\Delta V\,.
\end{equation}
The vacuum energy density $\rho^{\rm vacuum \;energy}_{\Delta V}$  in the volume $\Delta V$ is
\begin{equation}\label{eqn:46}
\rho^{\rm vacuum \; energy}_{\Delta V}=\frac{\Delta W}{\Delta\, V}= \rho^{\rm vacuum \; energy}\,,
\end{equation}
verifying that as the universe expands adiabatically\cite{Carroll:92,Carroll:01,Wilczek:08}, the vacuum energy density remains constant as it must since  the vacuum is everywhere homogeneous. 

\section{Summary and Discussion} 
\label{sec:5}
Since the action integral describing normal matter and energy is invariant under time translations, it follows from Noether's theorem\cite{Noether:18} that Normal energy is conserved.  Because the vacuum is homogeneous, the vacuum energy density must also be homogeneous.   That is, the expectation value of vacuum energy must be homogeneous.  If energy could be exchanged between normal energy and vacuum energy, normal energy would not be conserved and the vacuum energy density would not be homogeneous: vacuum energy and normal energy are conserved independently.  As a result, during the time that a vacuum fluctuation exists, any interaction between normal energy and vacuum energy must satisfy the condition that there is no net transfer of energy between normal and vacuum energy.  Therefore, vacuum energy does not contribute to the energy density of normal matter, explaining why there is no ``vacuum catastrophe''. It then follows that vacuum energy does not gravitate: if vacuum energy could create a gravitational field, that field could do work on normal energy, violating the independent conservation of normal energy and vacuum energy.  Consequently, the energy-momentum tensor resulting from vacuum energy should not be included in Einstein's field equation, resolving the ``old cosmological constant problem''.  

The term ``vacuum fluctuation''  is used by physicists to describe two different entities:  Type 1 vacuum fluctuations (VFs) are present for a time $\Delta t$ permitted by the uncertainty principle and are on-shell so they appear as external particles in a Feynman diagram.  VFs can interact with ordinary matter subject to the constraint that during the lifetime of each VF there is no net transfer of energy between vacuum and normal energy\footnote{The calculation of $\epsilon_0$ (from which formulas for $c$ and $\alpha$ immediately follow)  relies on the separate conservation of vacuum and normal energy.}. Type 2 vacuum fluctuations are a collection of interacting, virtual particles that arise from \textendash \;and then disappear back into \textendash \;the vacuum. Virtual particles are not actually particles because they only appear in perturbation calculations.  Since virtual particles are off-shell, they are represented by internal lines in a Feynman diagram, implying that type 2 vacuum fluctuations do not contribute directly to physical processes.

To minimize the violation of conservation of energy permitted by the uncertainty principle and to avoid violating conservation of angular momentum and conservation of quantum numbers such as charge etc., in an inertial frame VFs appear as particle-antiparticle pairs bound together in the lowest energy state that has zero angular momentum and a center of mass that is at rest. The VFs that contribute most to $\epsilon_0$ are bound states of charged lepton-antilepton pairs. When interacting with a photon, these bound states oscillate similarly to the way an atom or molecule in a dielectric oscillates: the calculation of the permittivity  $\epsilon_0$ of the vacuum is somewhat similar to the calculation of the permittivity of a dielectric. Using the formula for  $\epsilon_0$,   formulas for the speed of light in the vacuum and the fine-structure constant immediately follow, respectively, from $c=1/\sqrt{\mu_0\epsilon_0}$  and $\alpha \equiv e^2/(4 \pi \epsilon_0 \hbar c)$.  

The formula calculated here for the speed of light in the vacuum is  1.3\% less than the accepted value. It  is not more accurate because only the leading term has been calculated in what turns out to be an infinite series in $\alpha$. In addition to the fact that just the leading term in the series for $c$ almost agrees with the accepted value, there is another reason for thinking that the calculation is correct.  Since the speed of light in the vacuum is the same in any direction in every inertial frame, the value of of $c$ calculated here must satisfy that condition.  The permittivity of the vacuum $\epsilon_0$ is calculated by examining the electromagnetic interaction of photons with vacuum fluctuations.  Since both the vacuum and Maxwell's equations are the same in every inertial frame, the calculated value of $\epsilon_0$ is the same in every inertial frame. Because $c$ is calculated using $c=1/\sqrt{\epsilon_0 \mu_0}$, the calculated value of the speed of light in the vacuum is also the same in every inertial frame, as it must be. 

\section*{References}

\bibliographystyle{iopart-num}
\bibliography{2020}
\end{document}